\documentclass[preprint,12pt]{aastex}

\usepackage{floatflt, multicol}
\usepackage{graphicx}

\begin{document}

\title{X-ray Studies of Planetary Systems: \\
An Astro2010 Decadal Survey White Paper}

\author{Eric Feigelson$^1$, Jeremy Drake$^6$, Ronald Elsner$^2$,
  Alfred Glassgold$^3$, Manuel G\"udel$^4$, Thierry Montmerle$^5$,
  Takaya Ohashi$^7$, Randall Smith$^6$, Bradford Wargelin$^6$, and
  Scott Wolk$^6$} 
\affil{$^{1}$Pennsylvania State University, $^{2}$NASA's Marshall
  Space Flight Center, $^{3}$University of California, Berkeley,
  $^{4}$Swiss Federal Institute of Technology Z\"urich,
  $^{5}$Laboratoire d'Astrophysique de Grenoble,
  $^{6}$Harvard-Smithsonian Center for Astrophysics
  $^{7}$Tokyo Metropolitan University}
\maketitle

\pagebreak

\section{Introduction  \label{sec:planet_intro}}

\begin{floatingfigure}[r]{3in}
\noindent \includegraphics[totalheight=2in]{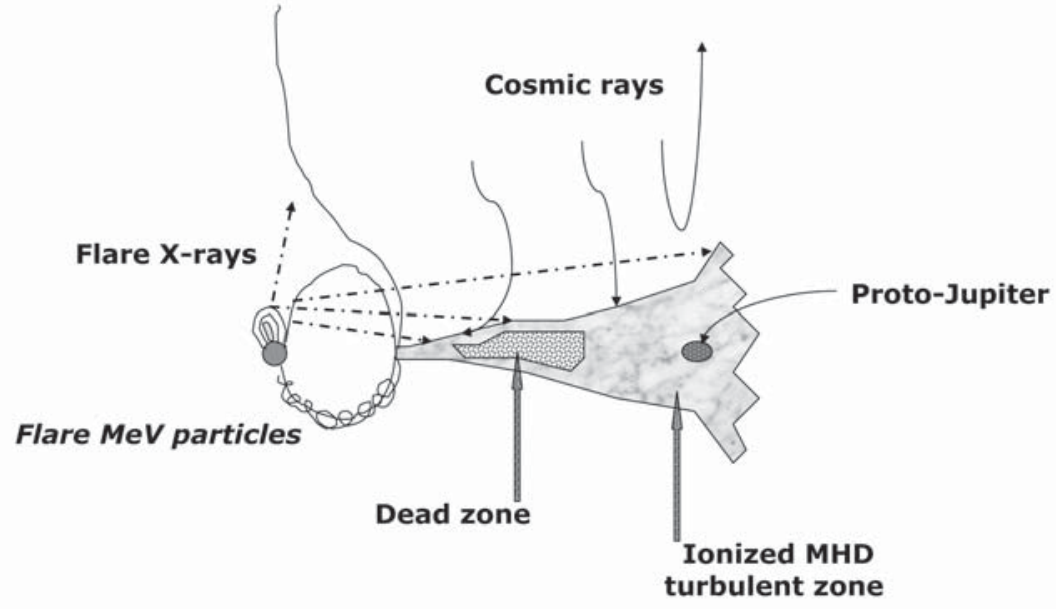}
\caption{\it Protoplanetary disk illuminated by flare X-rays and its
  effects on disk ionization.\label{fig:psd}}
\end{floatingfigure}

\noindent It may seem counterintuitive that X-ray astronomy should give any
insights into planetary systems: planets, and their natal
protoplanetary disks (PPDs), have temperatures which are far too cool
($100-1500$~K) to emit X-rays.  However, planets orbit stars whose
magnetized surfaces divert a small fraction of the stellar energy into
high energy products: coronal UV and X-rays, flare X-rays and
energetic particles, and a high-velocity stellar wind.  In our Solar
System, these components from the active Sun interact with the cool
orbiting bodies to produce X-rays through various processes including
charge-exchange between ionized and neutral components.  {\bf The
  resulting X-ray emission gives unique insights into the solar
  activity, planetary atmospheres, cometary comae, charge exchange
  physics, and space weather across the Solar System }\citep[review
  by][]{Bhardwaj07a}.

Solar-type stars also universally exhibit enhanced magnetic activity
during their youth.  X-ray emitting flares in pre-main sequence stars
are 100 - 10,000$\times$\ more powerful and frequent than in older
stars like today's Sun, and this emission only gradually declines over
the first billion years on the main sequence \citep[reviews
  by][]{Feigelson07, Guedel07a}.  The X-rays and energetic particles
from flares will irradiate protoplanetary disk gases and solids
\citep[review by][]{Glassgold00}.  As a result, it is possible that
      {\bf the stellar activity of young stars will substantially
        affect PPDs and planet formation processes}\ by heating disk
      outer layers, producing reactive ion-molecular species, inducing
      disk turbulence via ionization, and explaining conundrums in the
      meteoritic and cometary record such as isotopic anomalies,
      chondrule melting, and radial mixing (see Figure~\ref{fig:psd}).  Later, X-ray and
      ultraviolet irradiation will speed evaporation of planetary
      atmospheres and thereby perhaps affect planet habitability.

Discoveries of X-rays from Solar System bodies were made with the full
range of X-ray astronomical satellite observatories over the past
three decades, from the discovery of Jupiter's emission with the  {\it
  Einstein Observatory} and cometary comae with {\it ROSAT} to the
study today of planets and moons with the {\it Chandra X-ray
  Observatory} and {\it XMM-Newton}.   X-rays from young stars are now
investigated in thousands of pre-main sequence stars in the nearby
Galaxy.  Today, $\sim 40$ papers/year are published on the
observations and implications of X-ray emission relating to planetary
science.   But the X-ray emission is faint, time variable and
spectrally complex $-$ today's instrumentation can achieve only a
small portion of the potential scientific advances in planetary
sciences.  The planned high-throughput International X-ray Observatory
(IXO) will propel this nascent field forward. 

We highlight here five studies in planetary science done using X-ray
observations.  These studies address in unique ways several of NASA's
strategic goals \citep{SMD06} concerning the effects of the Sun on its
planets, the physics of planetary ionospheres and ion-neutral
interactions, the role of stellar activity on planet habitability, and
on the formation processes of planetary systems around young stars.
These studies will complement NASA's strong program on Solar System
exploration, extrasolar planet discovery, and planet formation
environments.

\section{Probing protoplanetary disks with the iron fluorescent
  line \label{sec:planet_disk}} 

\noindent The 19$^{th}$ century insights into the origin of our Solar System
involving gravitational collapse of cold gas with angular momentum
have been validated in recent decades by the profound discoveries of
infrared-emitting PPDs disks around nascent stars in nearby star
forming regions and discoveries of extrasolar planetary systems around
a significant fraction of older stars in the solar neighborhood.
However, a number of enigmatic phenomena have been noted which
indicate that non-equilibrium high energy processes play some role in
planet formation.  Laboratory study of meteorites and {\it Stardust}
cometary material, which record processes in the planetesimal stage of
our protoplanetary disk 4.567~Gyr ago, reveal flash-melted chondrules,
calcium-aluminum-rich inclusions and free-floating grains with
daughter products of short-lived spallogenic radionuclides, and
composites with annealed or glassy components \citep[reviews
  by][]{Connolly06, Chaussidon06}.  Infrared spectroscopic studies of
some distant PPDs with NASA's {\it Spitzer Space Telescope} reveal
heated and ionized gaseous outer layers \citep[reviews by][]{Najita07,
  Bergin07}.  While some of these phenomena can be attributed to the
effects of violent events which precede gravitational collapse (such
as supernova explosions), others require irradiation of disks by the
X-rays and energetic particles from magnetic reconnection flares
around the host young star.  X-ray ionization should dominate cosmic
ray ionization by several orders of magnitude
\citep{Glassgold00}. X-ray astronomers are thus joining the vibrant
community of meteoriticists, infrared and millimeter spectroscopists,
and theorists seeking {\bf to understand non-equilibrium processes during
the protoplanetary disk stage of planet formation.}

A particularly important consequence of X-ray irradiation of PPDs is
the predicted induction of MHD turbulence by coupling the
slightly-ionized gas to magnetic fields in a sheared Keplerian
velocity field via the magneto-rotational instability.  Harder X-rays
($>10$~keV) from powerful flares can penetrate deeply into
protoplanetary disks, and may reach the PPD midplane where planets
form.  Astrophysicists are enormously interested in the possibility of
turbulent PPDs as it appears to solve certain problems (e.g. gas
viscosity needed for accretion, inhibition of Type I migration of
larger protoplanets) while it raises other problems (e.g. inhibition
of grain settling to the disk midplane, promotion of shattering rather
than merger of small solid bodies).  Stellar X-rays may also be
responsible for the ionization needed to propel collimated
protostellar bipolar outflows, for the evaporation of icy mantles in
PPD grains, and for ion-molecular chemical reactions in the disks.
X-rays may play a critical role in the photoevaporation and
dissipation of older protoplanetary disks \citep{Ercolano08}.
Figure~\ref{fig:psd} illustrates various aspects of an X-ray
illuminated PPD.

\begin{figure}[t]
\includegraphics[totalheight=1.7in]{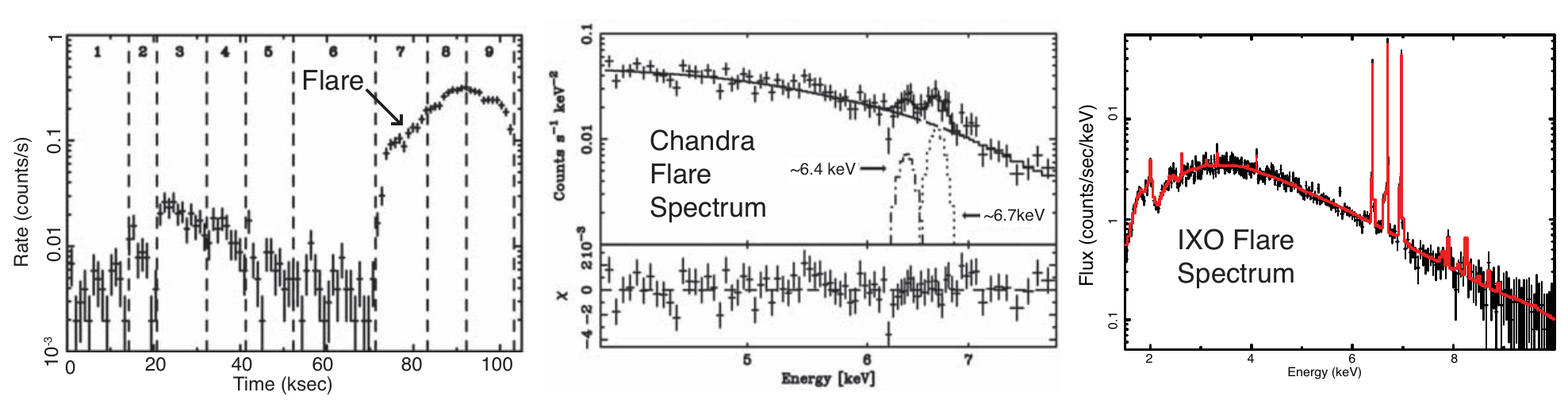}
\caption{\it Chandra light curve [Left] and CCD spectrum [Middle] showing the
Fe 6.4~keV fluorescent line during a powerful flare from the protostar
YLW~16A in the nearby Ophiuchi cloud \citep{Imanishi01}.  The
high-ionization emission lines (6.7 keV) arise from the hot plasma
confined in the flaring magnetic loop. [Right] Simulation of a 2~ks
IXO XMS spectrum at the onset of the YLW~16A flare showing the Fe
6.4~keV fluorescent line.  
IXO will be able to see flares 100x fainter than currently possible.
\label{fig:disk}}
\end{figure} 
The strongest test of X-ray irradiation of PPDs is the fluorescent
iron line at 6.4~keV, which is well-known to appear when a hard X-ray
continuum from a central source illuminates cool disk material (e.g.,
in enshrouded active galactic nuclei and X-ray binary systems).  The
6.4 keV emission line has been seen in a few flaring protostellar
systems (see Figure~\ref{fig:disk}) but typically lies beyond the
sensitivity limit of current instrumentation.  With the $\sim 200$
times improved sensitivity in the fluorescent line compared to {\it
  Chandra} and {\it XMM-Newton}, the IXO X-ray Microcalorimeter
Spectrometer (XMS) detector will detect (or place strong constraints
on) X-ray irradiation in hundreds of PPDs in the nearby Ophiuchus,
Taurus, Perseus and Orion star forming clouds.  The IXO Hard X-ray
Imager (HXI) will separately establish the intensity of
deeply-penetrating X-rays in the $10-30$ keV band needed to calculate
PPD turbulent and ``dead'' zones.  Prior to IXO's launch, the molecular
and dust properties of these disks will be well-characterized by the
{\it Spitzer}, {\it Herschel} and {\it James Webb} missions and ALMA
telescope.  By correlation of X-ray, molecular and solid properties in
these systems, IXO should clearly establish the role of X-ray
illumination on PPD physics and chemistry.  It is not impossible that
diversity in X-ray irradiation plays a critical role in the diversity
of exoplanetary systems seen around older stars.

\section{The complex X-ray emission of Jupiter and Mars  \label{sec:planet_Jup}}

\begin{floatingfigure}[r]{4.0in}
\includegraphics[totalheight=2.0in]{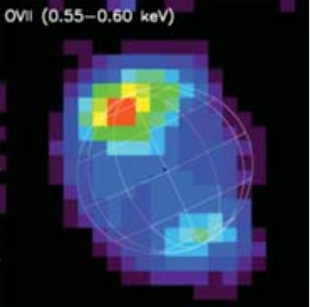}
\includegraphics[totalheight=2.0in]{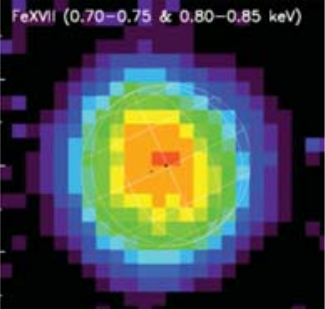}
\caption{ {\it XMM-Newton images of Jupiter in the charge-exchange
  O~VII [left] and fluorescent Fe XVII [right] lines.}\label{fig:JupiterImage}}
\end{floatingfigure}

\noindent Jupiter is the most luminous X-ray emitter in the Solar System after
the Sun.  Its emission is complex with several spatially and
temporally varying components: charge exchange lines from interaction
with solar wind ions, fluorescence and scattering of solar X-rays, and
a hard electron brems-strahlung emission \citep{Bhardwaj07a}.  Charge
exchange is the dominant process where heavy solar energetic ions
collide with neutral atoms in the planetary atmosphere, producing a
radiative cascade of non-thermal emission lines (e.g. from the $n=5$
state of hydrogenic O$^{7+}$ at 0.653~keV).  In planets like Earth and
Jupiter with a strong dipolar magnetic field, these X-ray components
are concentrated in auroral regions around the north and south
magnetic poles (see Figure~\ref{fig:JupiterImage}).  K-shell
fluorescence from carbon and oxygen is the dominant X-ray emitting
process from Venus and Mars where the atmospheres are rich in CO$_2$
and wind ions are not concentrated toward the poles by strong magnetic
fields.  High-amplitude variations on timescales of minutes-to-hours
can be present in these X-ray components.

\begin{figure}[t]
\includegraphics[totalheight=2.0in]{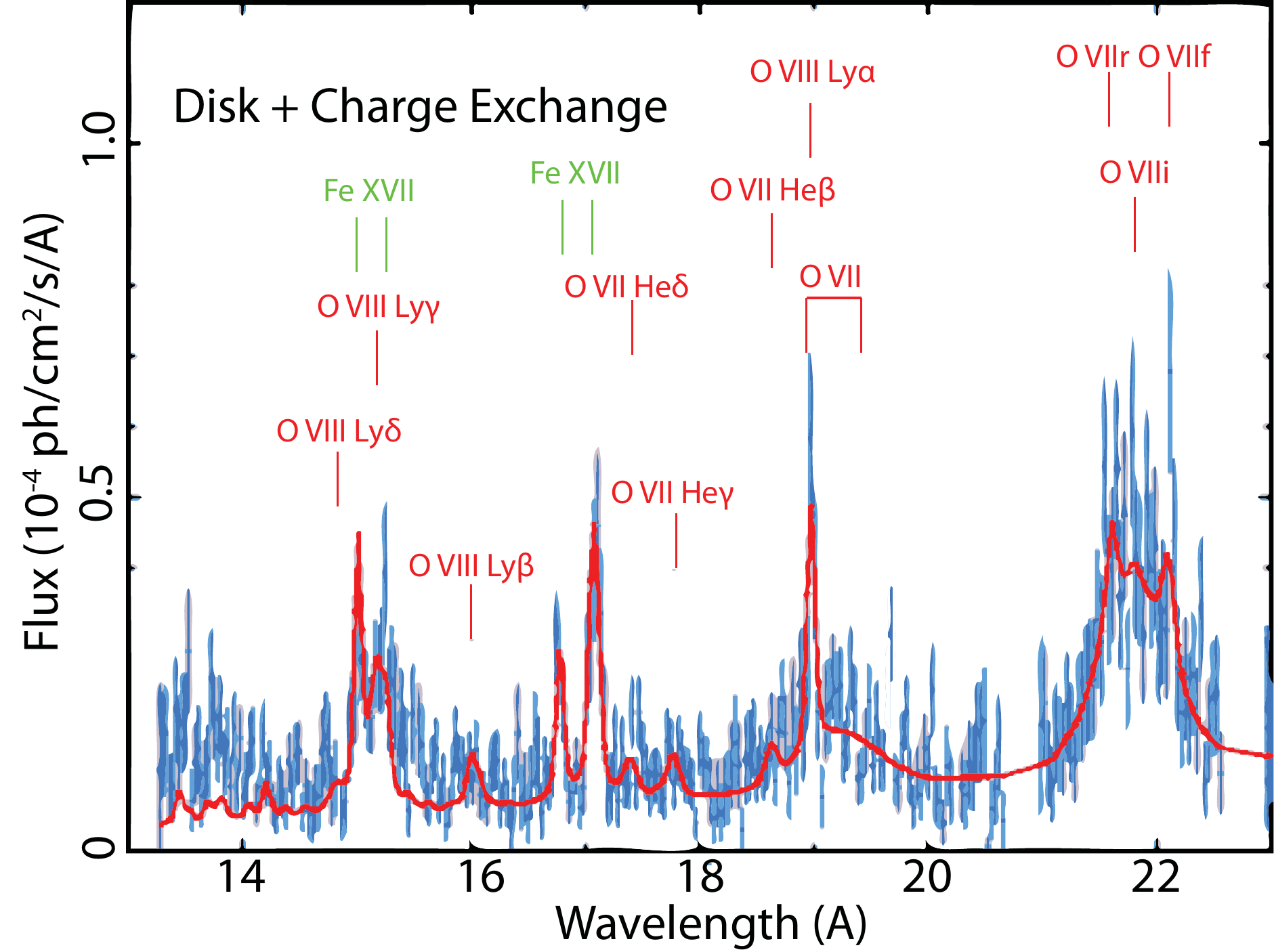} \hskip 0.2in
\includegraphics[totalheight=2.0in]{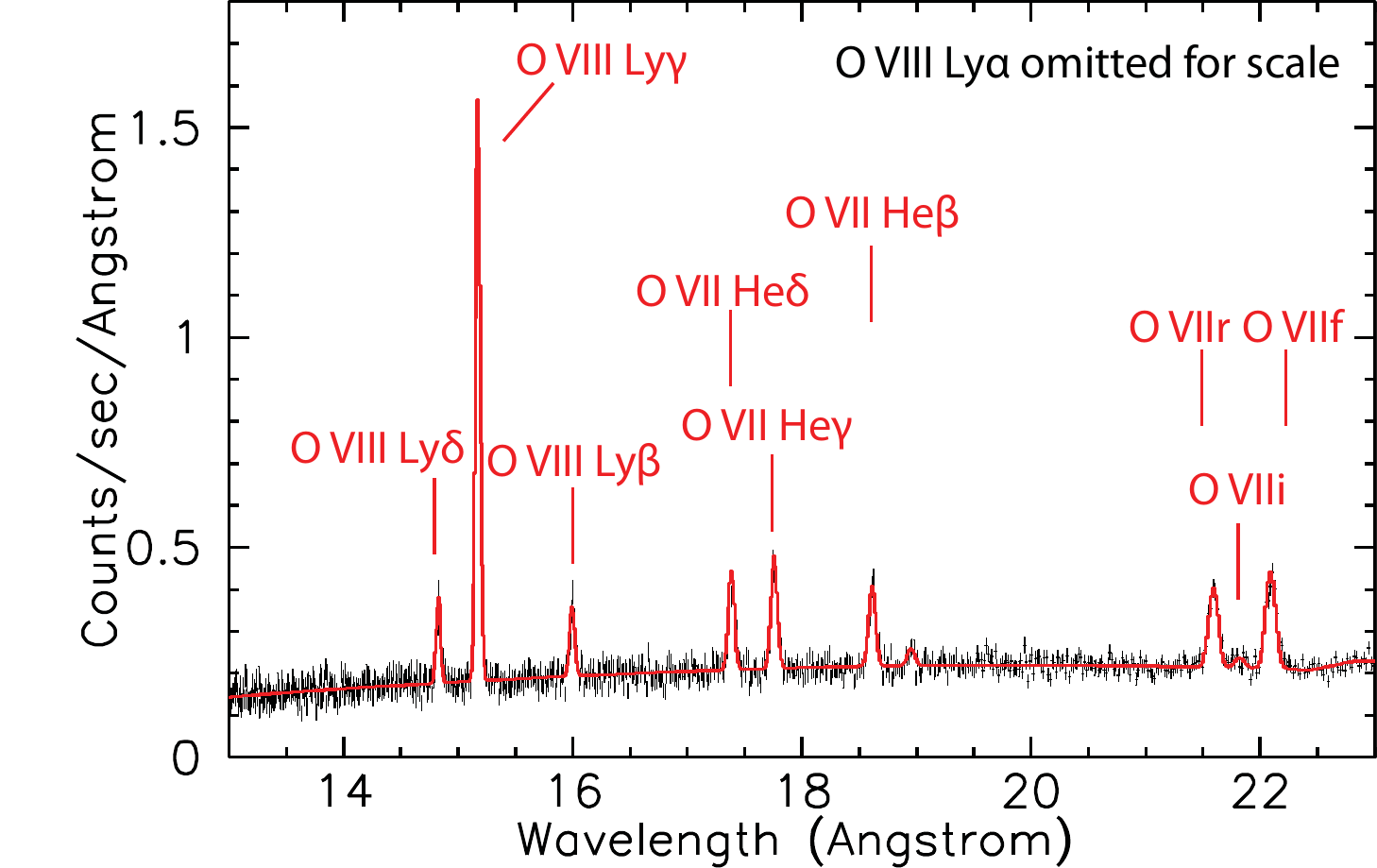} 
\caption{[Left] {\it XMM-Newton spectrum of Jupiter (blue) and a
    three-component model (red) consisting of auroral charge-exchange
    lines, solar reflection continuum, hard electron bremsstrahlung
    continuum\citep{Branduardi07}.  [Right] Simulation of a 50~ks IXO
    spectrum based on the XMM-Newton emission model showing the
      oxygen charge-exchange emission lines; many more lines are
      expected.}  \label{fig:Jupiter} }
\end{figure}

Figure \ref{fig:Jupiter} shows the best spectra from Jupiter currently
available.  The IXO spectrum will reveal hundreds of lines with
sufficient signal to map the upper atmosphere through the planet's
36~ks rotational period.  Repeated visits, particularly at different
periods in the solar 11-year activity cycle and several days after
powerful solar flares, should show varying ratios of the different
emission components elucidating the complex physics of solar-planetary
interactions.   

An IXO study of Mars may be particularly important for understanding the
evaporative effect of solar X-rays and extreme ultraviolet emission on
planetary atmospheres.  Martian X-ray emission is dominated by a
uniform disk of scattered solar radiation (Mars subtends 18\arcsec\/
at opposition).  But remarkably, a faint halo of soft charge exchange
lines with unexplained spatial substructure is seen out to $\sim 8$
planetary radii \citep[Figure~\ref{fig:Mars},][]{Dennerl02}.  This
exceedingly faint X-ray component gives a unique view into planetary
exospheres which is inaccessible at other wavelengths.  When observed
with IXO under a variety of solar wind and flare conditions, the
Martian exosphere may provide critical evidence into the complex
interactions between stellar X-ray and ultraviolet emission and
planetary atmospheres.  Indeed, it is possible that Mars' atmosphere
is so thin today due to these effect when solar magnetic activity was
greatly elevated $\sim 4$~Gyr ago.

\section{Cometary charge exchange} 

\begin{floatingfigure}[r]{2.5in}
\includegraphics[totalheight=1.9in]{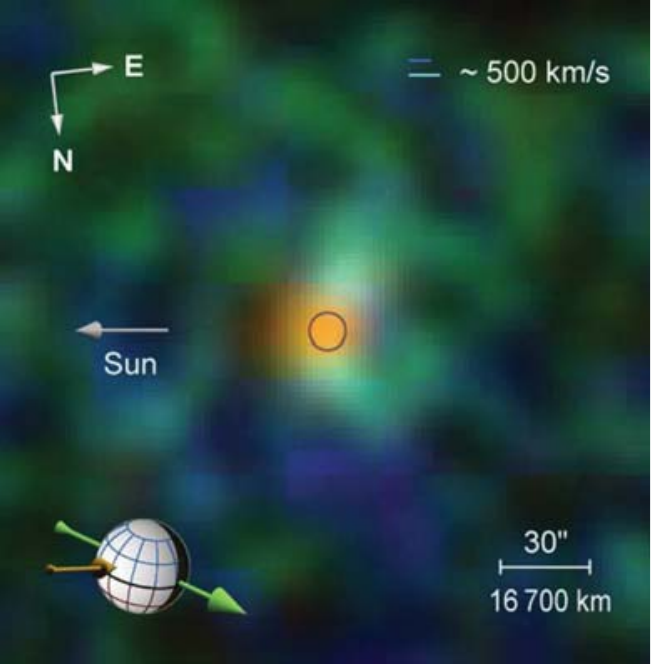}
\caption{\it Image of Mars in X-rays derived from {\it XMM-Newton}
  observations: charge exchange O$^{+6}-$O$^{+7}$ lines in blue,
  C$^{+4}-$C$^{+5}$ lines in green, and fluorescent lines in yellow
  \citep{Dennerl06}. \label{fig:Mars}} 
\end{floatingfigure}

\noindent During their perihelion approach to the Sun, cometary ices (mostly
water) are evaporated into a large neutral coma which produces strong
charge exchange reactions when it interacts with highly-charged solar
wind ions \citep{Cravens97}.  X-ray studies provide unique information
on this wind-coma interaction region giving insights into charge
exchange processes, wind-coma hydrodynamics, and cometary outgassing.
Interpretation of cometary X-ray spectra today is complicated as it
depends on the solar wind velocity, density and composition, as well
as wind ion penetration into the coma, ion-molecular cross-sections,
and collisional opacity.  These issues can be elucidated by IXO XMS
studies (note that grating observations are not feasible due to the
spatial extent).  X-ray luminosities range from $10^{14} - 10^{16}$
erg/s depending primarily on the comets' encounters with different
solar wind states \citep{Bodewits07}.  Several periodic comets and an
unknown number of distant comets will enter perihelion during the IXO
mission.  Spectra will resolve about a dozen charge-exchange emission
lines from oxygen above 0.5~keV, and dozens of lines from other
elements.  Line ratios will change with cometary gas species, solar
wind ion composition and wind speed.

\section{Heliospheric charge exchange} 

\noindent X-ray astronomers have increasingly recognized that a significant
fraction of the all-sky soft X-ray background arises from
time-dependent heliospheric charge-exchange reactions between highly
ionized solar wind atoms and interstellar neutrals which penetrate
deeply into the heliosphere \citep{Snowden04}.  This has profound
implications for our understanding of the Local Hot Bubble and the
structure of the Galactic interstellar medium.   Earlier X-ray
missions also suffer from contamination by charge exchange emission
within the terrestrial magnetosphere \citep{Wargelin04}, but IXO will
avoid this component from its location at the Earth-Sun L2 Lagrangian
point. From study of the background of dozens of observations with
different lines-of-sight through the heliosphere under different solar
wind conditions, IXO spectra should provide powerful insights into
heliospheric physics and its interactions with its ambient Galactic
medium.

\section{Atmosphere evaporation in extrasolar planets} 

\noindent Past studies of solar-type stars indicate that X-ray luminosities drop
roughly 10-fold between ages of $10^7$ and $10^8$~yr, another 10-fold
between $10^8$ and $10^9$~yr, and more rapidly between $10^9$ and
$10^{10}$~yr \citep{Preibisch05}.  This enhanced X-ray emission during
early epochs, and the associated extreme ultraviolet emission which is
more difficult to study in young stars, will dissociate and ionize
molecules in planetary thermospheres and exospheres so that light
atoms escape into the interplanetary medium \citep{Guedel07a, Penz08}.  Solar
wind and flare particles may also erode the {\it entire}\ atmosphere
if no magnetic field is present.  These processes were probably
important on Venus, Earth and Mars during the first $10^8$~yr and are
presently leading to hydrodynamics escape of the atmospheres in
extrasolar ``hot Jupiters.''  The atmospheric conditions, and hence the
habitability, of planets may thus be regulated in part by the
evolution of the ultraviolet and X-ray emission of their host stars.
Thousands of extrasolar planets will be known by 2020 through NASA's
{\it Kepler} mission and other planetary search programs.  IXO can
measure both the quiescent and flare activity of specific stars which
will be known to have planets in their habitable zones.  Combined with
stellar activity evolutionary trends and planetary atmospheric
modeling, IXO findings should give unique insights into the
atmospheric history of these potentially habitable planets.

\section{Summary}

\noindent X-ray studies of planetary systems are beginning to provide important
insights into planetary astrophysics which are inaccessible at other
wavelengths.  X-ray observations of the host stars reveal the
high-energy inputs to protoplanetary disks and planetary atmospheres
due to stellar winds and violent magnetic flaring. X-rays from Solar
System planets are faint but reveal considerable complexity, a
situation well-matched to IXO's high-throughput and high spectral
resolution.  Charge exchange line emission from interactions between
solar wind ions and atmospheric neutrals, along with other processes,
are seen in the atmospheres of Jupiter, comets and other Solar System
bodies.  The X-ray discovery of the Martian exosphere points to
evaporation of planetary atmospheres unprotected by magnetic fields,
which may play an important role in the habitability of planets.
X-ray and infrared spectroscopic studies of protoplanetary disks show
that X-ray ionization is present, and theoretical calculations
indicate its importance to disk thermodynamics, chemistry and
dynamics.  It is possible that X-ray illumination is a critical
regulator to the formation and early evolution of planets in the disk,
but higher sensitivity is needed to study the crucial 6.4~keV
fluorescent line.   X-ray studies of cometary coma charge exchange,
charge exchange distributed throughout the heliosphere, and of stars
hosting extrasolar planets are examples of a wealth of IXO studies
which will revolutionize this field.

\end{document}